\DocumentMetadata{}
\documentclass[sigconf,10pt]{acmart}
\usepackage{epsfig,endnotes}
\usepackage{amsmath}
\usepackage{algorithmic}
\usepackage[linesnumbered,ruled]{algorithm2e}
\usepackage{mathtools}
\usepackage{graphicx}
\usepackage[capitalize]{cleveref}
\usepackage{xspace}
\usepackage{caption}
\usepackage{textcomp}
\usepackage{comment}
\usepackage{booktabs}
\usepackage{subcaption}
\usepackage{fancyhdr}
\usepackage{enumitem}

\newenvironment{denseitemize}{
\begin{itemize}[topsep=2pt, partopsep=0pt, leftmargin=1.5em]
  \setlength{\itemsep}{2pt}
  \setlength{\parskip}{0pt}
  \setlength{\parsep}{0pt}
}{\end{itemize}}

\newenvironment{denseenum}{
\begin{enumerate}[topsep=2pt, partopsep=0pt, leftmargin=1.5em]
  \setlength{\itemsep}{2pt}
  \setlength{\parskip}{0pt}
  \setlength{\parsep}{0pt}
}{\end{enumerate}}

\def\eg{{e.g.,\ }}
\def\etal{{et al.\xspace}}

\usepackage{tikz}

\usepackage{framed}

\newif\ifxetexorluatex
\ifxetex
  \xetexorluatextrue
\else
  \ifluatex
    \xetexorluatextrue
  \else
    \xetexorluatexfalse
  \fi
\fi
\ifxetexorluatex%
  \usepackage{fontspec}
  \usepackage{libertine} 
  \newfontfamily\quotefont[Ligatures=TeX]{Linux Libertine O} 
\else
  \usepackage[utf8]{inputenc}
  \usepackage[T1]{fontenc}
  \usepackage{libertine} 
  \newcommand*\quotefont{\fontfamily{LinuxLibertineT-LF}} 
\fi

\newcommand*\quotesize{30} 
\newcommand*{\openquote}
   {\tikz[remember picture,overlay,xshift=-4ex,yshift=-2.5ex]
   \node (OQ) {\quotefont\fontsize{\quotesize}{\quotesize}\selectfont``};\kern0pt}

\newcommand*{\closequote}[1]
  {\tikz[remember picture,overlay,xshift=4ex,yshift={#1}]
   \node (CQ) {\quotefont\fontsize{\quotesize}{\quotesize}\selectfont''};}

\colorlet{shadecolor}{white}

\newcommand*\shadedauthorformat{\emph} 

\newcommand*\authoralign[1]{%
  \if#1l
    \def\authorfill{}\def\quotefill{\hfill}
  \else
    \if#1r
      \def\authorfill{\hfill}\def\quotefill{}
    \else
      \if#1c
        \gdef\authorfill{\hfill}\def\quotefill{\hfill}
      \else\typeout{Invalid option}
      \fi
    \fi
  \fi}
{\authoralign{#1}
\ifblank{#2}
   {\def\shadequoteauthor{}\def\yshift{-2ex}\def\quotefill{\hfill}}
   {\def\shadequoteauthor{\par\authorfill\shadedauthorformat{#2}}\def\yshift{2ex}}
\begin{snugshade}\begin{quote}\openquote}
{\shadequoteauthor\quotefill\closequote{\yshift}\end{quote}\end{snugshade}}

\acmYear{2025}\copyrightyear{2025}
\setcopyright{rightsretained}
\acmConference[HOTOS 25]{Workshop in Hot Topics in Operating Systems}{May 14--16, 2025}{Banff, AB, Canada}
\acmBooktitle{Workshop in Hot Topics in Operating Systems (HOTOS 25), May 14--16, 2025, Banff, AB, Canada}
\acmDOI{10.1145/3713082.3730393}
\acmISBN{979-8-4007-1475-7/25/05}

\begin{document}

\title{My CXL Pool Obviates Your PCIe Switch}

\newcommand*{\microsoft}{\includegraphics[width=1em]{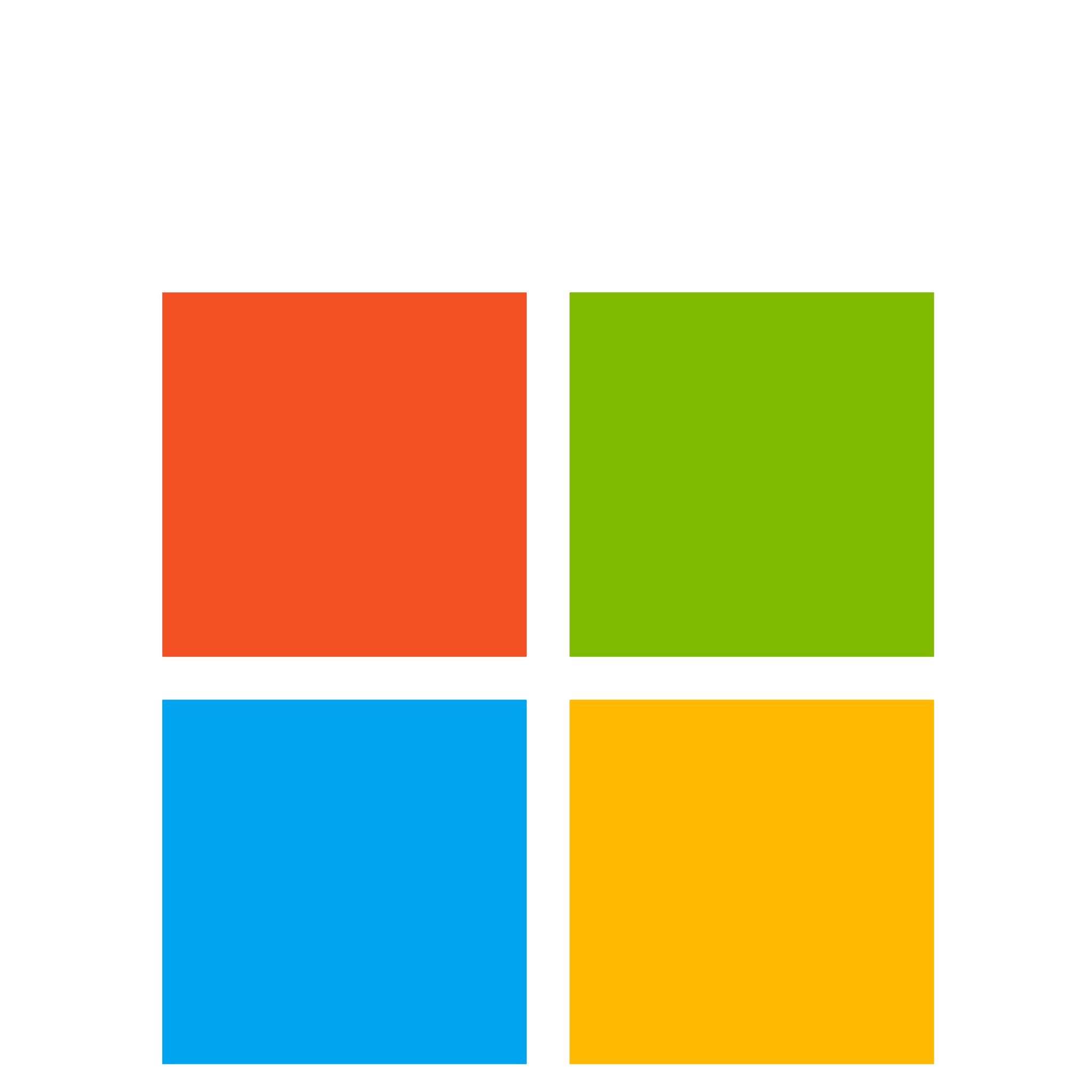}}%
\newcommand*{\uw}{\includegraphics[width=1em]{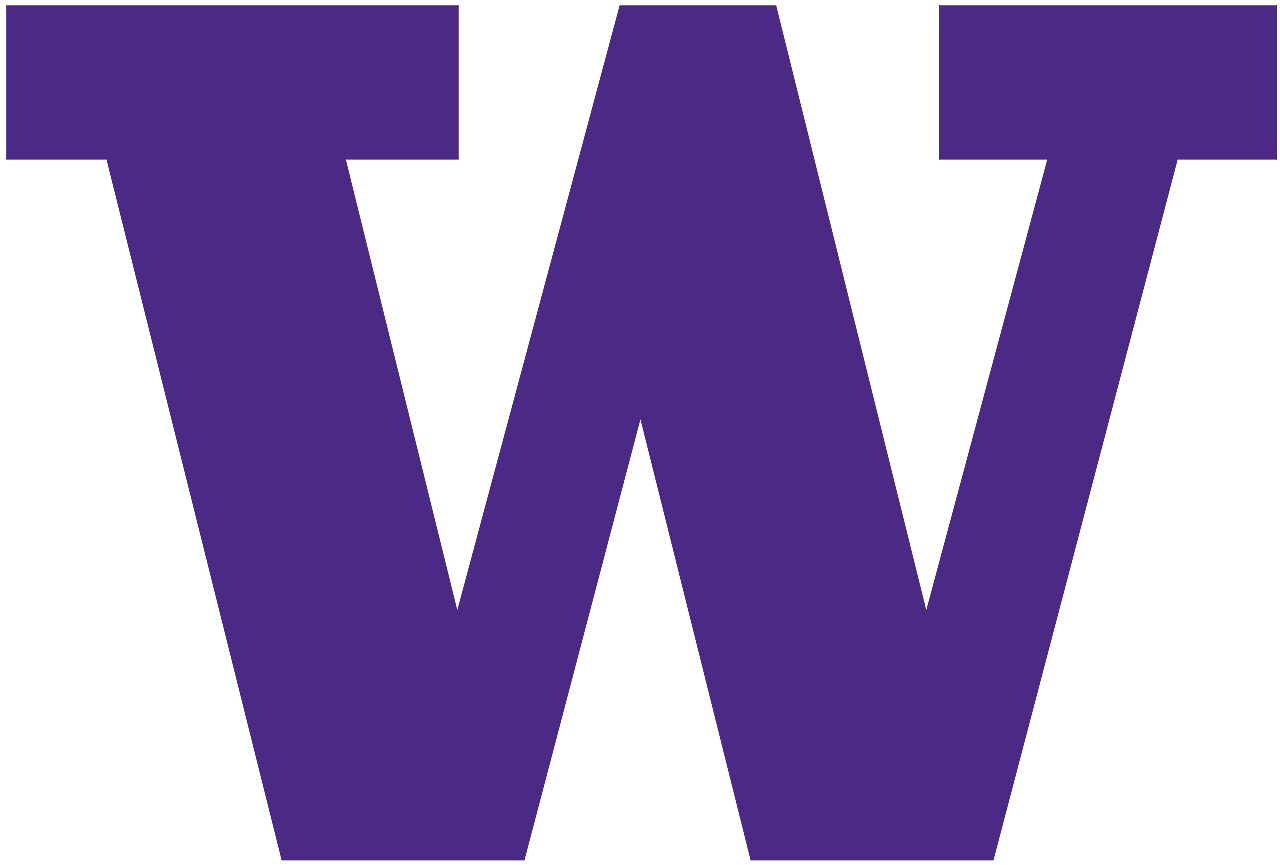}}%
\newcommand*{\columbia}{\includegraphics[width=1em]{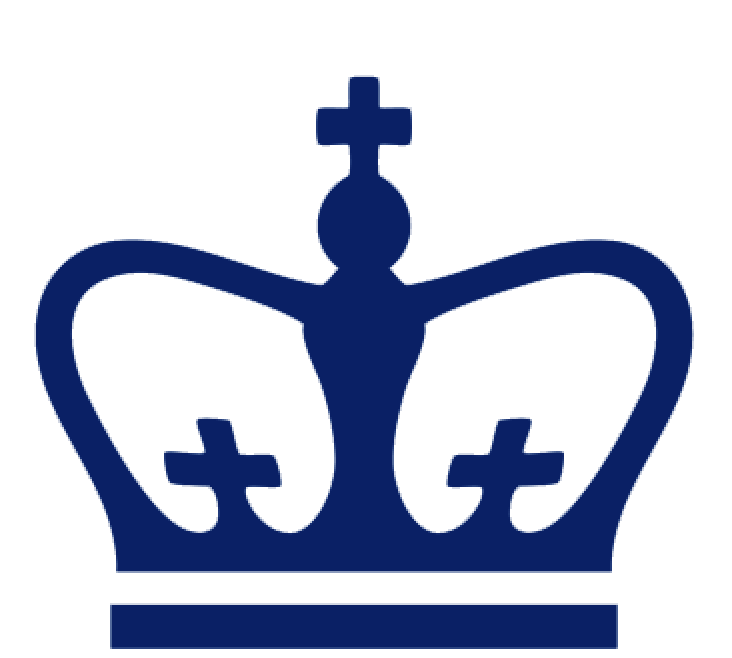}}%

\author{Yuhong Zhong}
\affiliation{%
    \institution{Columbia University}
}

\author{Daniel S. Berger}
\affiliation{%
    \institution{Microsoft Azure}
    \institution{University of Washington}
}

\author{Pantea Zardoshti}
\affiliation{%
    \institution{Microsoft Azure}
}

\author{Enrique Saurez}
\affiliation{%
    \institution{Microsoft Azure}
}

\author{Jacob Nelson}
\affiliation{%
    \institution{Microsoft Research}
}

\author{Antonis Psistakis}
\affiliation{%
    \institution{University of Illinois Urbana-Champaign}
}

\author{Joshua Fried}
\affiliation{%
    \institution{MIT CSAIL}
}

\author{Asaf Cidon}
\affiliation{%
    \institution{Columbia University}
}

\renewcommand{\shortauthors}{Zhong \etal}
\pagenumbering{gobble}

\begin{abstract}

Pooling PCIe devices across multiple hosts offers a promising solution to mitigate stranded I/O resources, enhance device utilization, address device failures, and reduce total cost of ownership.
The only viable option today are PCIe switches, which decouple PCIe devices from hosts by connecting them through a hardware switch.
However, the high cost and limited flexibility of PCIe switches hinder their widespread adoption beyond specialized datacenter use cases.

This paper argues that PCIe device pooling can be effectively implemented in software using CXL memory pools.
CXL memory pools improve memory utilization and already have positive return on investment.
We find that, once CXL pools are in place, they can serve as a building block for pooling any kind of PCIe device.
We demonstrate that PCIe devices can directly use CXL memory as I/O buffers without device modifications, which enables routing PCIe traffic through CXL pool memory.
This software-based approach is deployable on today's hardware and is more flexible than hardware PCIe switches.
In particular, we explore how disaggregating devices such as NICs can transform datacenter infrastructure.

\end{abstract}


%

\maketitle
\section{Introduction}

PCIe devices --- such as NICs, SSDs, and accelerators --- account for a significant portion of server cost and power consumption~\cite{greensku,dc-book,server-costs}.
Even optimized platforms like AWS and Azure use servers that physically connect a dozen SSDs over PCIe~\cite{aws2025instancestore, lyu2023hyrax}.
All servers use at least one high-bandwidth NIC.
This tight coupling of PCIe devices to servers requires providers to over-provision PCIe devices for peak demands and to offer redundancy for handling device failures. This leads to low utilization of NICs~\cite{poutievski2022jupiter, coach} and local SSDs~\cite{narayanan2016ssd, coach}.
Unused I/O resources are stranded within a host and cannot be used for I/O-intensive workloads running on other hosts.
A common approach to address resource stranding is to pool resources \emph{across multiple hosts}~\cite{pond,memtrade,blockflex,harvest-vm}.
Pooling resources within half a rack can already significantly improve utilization~\cite{pond,ahn2024examination}.

PCIe pooling enables multiple hosts to use any PCIe device in a shared pool. This would bring multiple benefits:
\begin{denseitemize}
    \item \textbf{Utilization.} Pooling PCIe devices mitigates stranded I/O resources and improve device utilization, which
    reduces datacenter total cost of ownership (TCO)~\cite{pond, coach}.
    \item \textbf{Failover.} When a PCIe-attached device fails, the host using it can fail over to other devices in the pool automatically. Pooling also reduces the number of redundant devices required for fault tolerance in a rack.
    \item \textbf{Load Balancing.} To prevent high load and high latency from PCIe device saturation, pools can dynamically adjust the number of hosts using a PCIe device by migrating workloads to less-utilized devices.
    \item \textbf{Peak Performance.} During demand spikes, a host can harvest all the PCIe devices in the pool to achieve higher aggregated performance~\cite{dfabric}.
    \item \textbf{Enabling Hardware Innovation.} When new accelerators (\eg smart SSDs and FPGAs) are deployed in public clouds, they often face low utilization. Pooling addresses this by allowing cloud providers to deploy a small number of accelerators (e.g., 1:16 ratio) while ensuring all hosts in the target racks can access them.
\end{denseitemize}

However, practical solutions for pooling PCIe devices have been limited and PCIe switches are the only viable option today.
One might think to use RDMA, since cloud providers already utilize RDMA to disaggregate SSDs~\cite{li2023more,nvmeof,azure-rdma}.
However, in practice, RDMA latency is too high; all cloud providers still offer host-local SSDs in addition to remote SSDs.
Anecdotally, AWS has tried to remove local SSDs from their offers, but brought them back due to customer demand\footnote{E.g., M6i supported local NVMe drives, M7i did not support local NVMe drives and was introduced in 2023~\cite{awsEC2M7i}. AWS later introduced I7ie~\cite{awsEC2I7ie} with local NVMe drives.}.
Pooling the NICs themselves over RDMA is unfeasible.

\begin{figure}
    \centering
    \includegraphics[width=0.8\columnwidth]{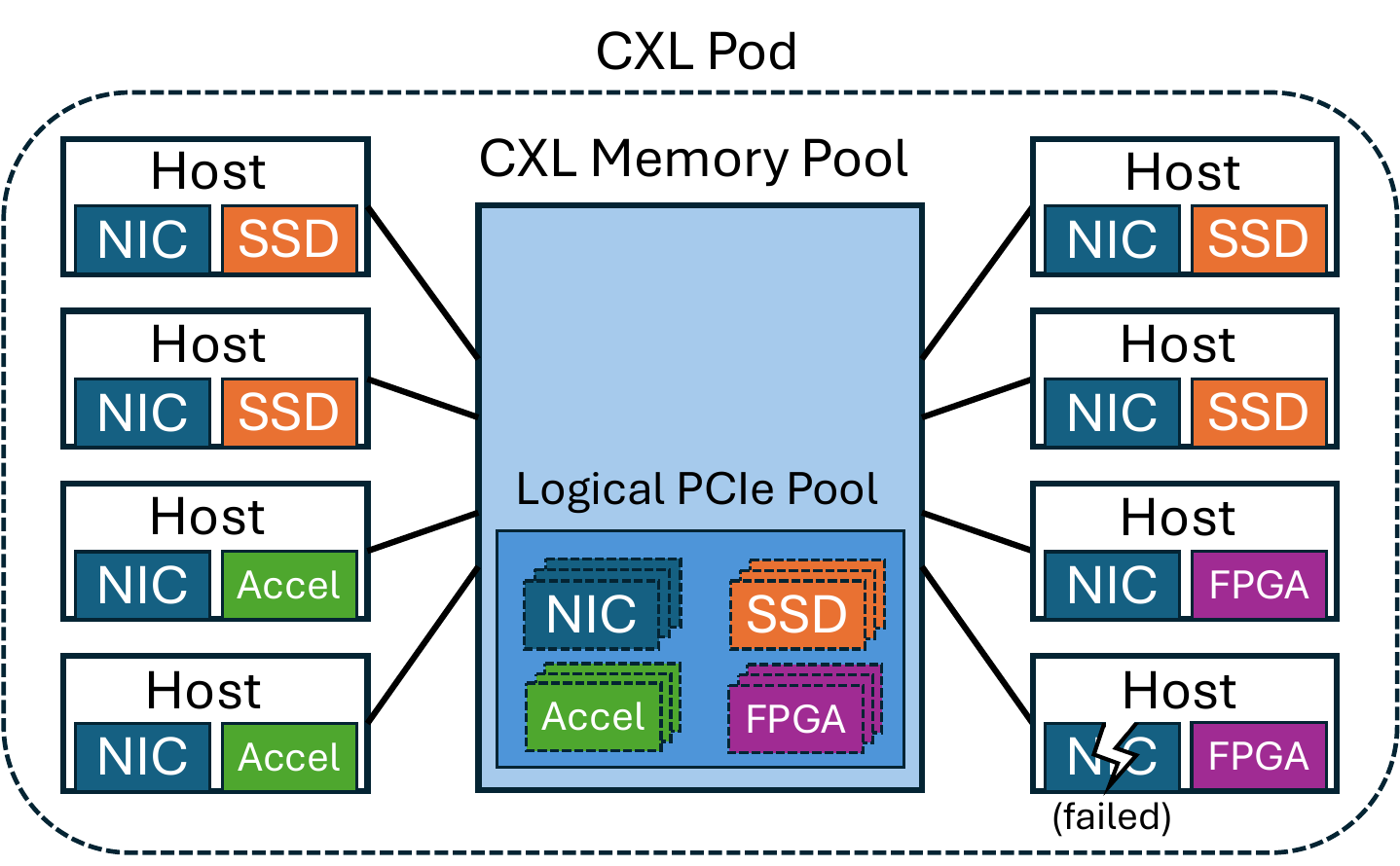}
    \vspace*{-3mm}
    \caption{A CXL memory pool enables hosts to access any PCIe device in a CXL pod, forming a logical pool of PCIe devices. }
    \label{fig:pcie-pool}
\end{figure}

PCIe switches have been \emph{the only} generic solution to pool PCIe devices~\cite{rpcie-bench,gigaio-news}.
Hosts and PCIe devices connect to a common PCIe switch which allows any host to utilize any device in the pool~\cite{gigaio-dev,liquid-dev,h3-dev,microchip-dev}.

Despite being technically available, deploying PCIe switches is costly and inflexible~\cite{rpcie-bench}.
The total cost of using PCIe switches in a rack, including the expenses for PCIe switches, switch software, host adapter cards, and cabling, easily reaches \$80,000~\cite{gigaio-price}.
Realistic deployments require redundant switches for fault tolerance and firmware updates, further increasing costs.
Such high costs can easily outweigh the cost savings of pooling. 
Additionally, as a hardware solution, PCIe switches are inflexible.
Different vendors have varying requirements for the topology between hosts and PCIe devices, as well as the types of PCIe devices they can support~\cite{liquid-dev,h3-dev,gigaio-dev}.
For example, Liquid's SmartStack is specifically designed to pool GPUs and does not support other device types~\cite{liquid-dev}, and GigaIO's FabreX has separate pooling appliances for storage devices and accelerators~\cite{gigaio-acc,gigaio-storage}.

In this paper, we argue that \emph{PCIe device pooling can be implemented in software} on top of Compute eXpress Link (CXL) memory pools.
CXL memory pools have gained industry interest to improve memory utilization and scale up in-memory databases~\cite{ahn2024examination,pond,hyperscale2023cxl,seagate2024cxl,ha2023dynamic,marvell2024structera,ufx-news,astera-news,cxl-db,lerner2024cxl}.
Hardware vendors have also started offering CXL memory with pooling capabilities\footnote{Although advanced pooling features (\eg Dynamic Capacity Device (DCD), Back Invalidation (BI)) are only standardized in the recent CXL 2.0 and 3.0 specification~\cite{cxl-3.0,cxl-2.0}, the widely adopted CXL 1.1 specification already provides the necessary functionalities for CXL memory pools~(\S\ref{sec:cxl})}~\cite{ufx-news,astera-news,marvell2024structera}.
A CXL pod~\cite{witchel2024challenges} consists of multiple hosts within a rack, allowing these hosts to dynamically allocate memory from the pool based on demands.
Recent work shows how to build CXL pods with hardware available today for about $\$600$ per host~\cite{berger2025octopus} \emph{without} using expensive CXL switches~(\S\ref{sec:cxl}).
This enables positive return-on-investment (ROI) for memory-pooling use cases~\cite{berger2023design}.

Once CXL pods are deployed, they can do much more than pooling memory --- they can serve as a building block for implementing PCIe device pooling in software (Figure~\ref{fig:pcie-pool}).
PCIe devices and hosts access CXL memory just like any other memory.
So, devices can directly read from or write to I/O buffers on CXL memory.
CPUs from multiple hosts can transparently access these I/O buffers. 


Implementing an efficient data path for PCIe device pooling in software brings multiple challenges.
First, CXL increases access latency by 2-3$\times$ compared to local DDR5 memory~\cite{cxl-uiuc,cxl-intro}.
We need to carefully assess the performance impact of placing I/O datapaths in CXL memory.
Second, although the CXL 3.0 specification introduces hardware coherence flows across hosts~\cite{cxl-3.0}, CXL memory pool devices available today are not cache-coherent across hosts.
We must therefore implement our own software coherence.

We argue that these challenges can be overcome with hardware available today.
A small fraction of memory from the CXL pool serves as (software-coherent) shared memory accessible to multiple hosts.
We show that a PCIe device attached to one CPU can DMA to/from shared memory
and that an application on another CPU --- to which the device is not connected via PCIe --- can access IO buffer data in shared memory.
Latency and bandwidth overheads are within 5\%.


To enable remote hosts to access PCIe device memory (\eg send MMIO commands or ring MMIO doorbells), we implement a sub-$\mu$s-scale host-to-host communication mechanism based on CXL shared memory.
To the best of our knowledge, this is the first paper to demonstrate shared-memory communication latency using a real CXL memory pool. Prior work assumes cross-host cache coherence and \emph{emulates} CXL pools using local memory~\cite{rpcool,hydrarpc,cxl-shm}.
This building block is used to forward device memory operations from remote hosts to the \emph{local host} where the devices are physically attached.
Our software design of PCIe device pooling offers unique benefits in terms of costs and flexibility:
First, CXL memory pools are inherently cheaper than PCIe switches due to switch-less pod designs~\cite{berger2025octopus}.
Second, PCIe pooling reuses the exact same hardware used for memory pooling use cases.
We can essentially enable PCIe pooling at no extra cost once CXL memory pools are deployed. 
Third, this software-based approach can dynamically support dynamic assignment of PCIe devices to hosts including potential fail-over scenarios.
Fourth, the software solution can be easily extended to support new PCIe devices without hardware changes.

In this paper, we first motivate PCIe device pooling (\S\ref{sec:motivation}) and provide background on CXL memory pools (\S\ref{sec:cxl}).
We then sketch our design of software-based PCIe pooling with CXL (\S\ref{sec:design}).
Lastly, we discuss how PCIe device pooling will change the datacenter infrastructure and other open questions, including how the CXL link bandwidth affects different PCIe pooling use cases (\S\ref{sec:discussion}).

\section{Why is PCIe Pooling a Good Idea?}
\label{sec:motivation}

\subsection{Stranded I/O Resources}

\begin{figure}
    \centering
    \includegraphics[width=0.6\columnwidth]{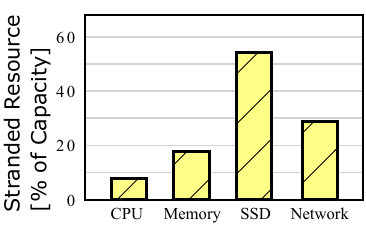}
    \vspace*{-3mm}
    \caption{Percentages of stranded CPU cores, memory capacity, SSD storage, and NIC bandwidth in Microsoft Azure datacenters.}
    \label{fig:stranding}
\end{figure}

Without pooling, PCIe devices are only accessible to a single host.
When a host depletes one type of resource (\eg CPU cores, memory, SSD storage, or NIC bandwidth), the remaining unutilized I/O resources are effectively \emph{stranded}, as additional workloads cannot be scheduled.

An key contributor to stranding is the heterogeneity of workloads, e.g., of Virtual Machine (VM) types in public clouds~\cite{ec2,gcp-vm,azure-vm}, which leads to a multi-dimensional bin-packing problem~\cite{legoos,tetris,protean}.
A host accepts new VMs until it fills up along one dimension (e.g., memory), leaving the other dimensions stranded (e.g., SSD capacity or NIC bandwidth).
For example, Microsoft Azure reports stranding across multiple resources (Figure~\ref{fig:stranding})~\cite{coach}.
We find that SSD capacity and NIC bandwidth are the two most stranded resources with 54\% and 29\% being stranded on average, respectively.

By pooling resources among $N$ servers, the effective bin’s shape becomes more flexible, so fewer resources become fully saturated in one dimension while sitting idle in others.
If demands across servers are somewhat independent, a spike in one server’s demand for a resource (e.g., lots of SSD usage) may be offset by lower demand in another server.
So, in aggregate, fewer resources remain stranded.

As a rough estimate, queueing theory typically shows a square-root improvement in resource overprovisioning when demands are aggregated over $N$ hosts.
Specifically, if demands across servers were independent, then the fraction of stranded resources would decrease with $\sqrt{N}$~\cite{Whitt1992,Janssen2011}.
So, pooling across even just $N=8$ servers would reduce SSD stranding from 54\% to 19\% and NIC stranding from 29\% to 10\% in Microsoft Azure.
If scheduling dependencies (e.g., availability zones and other constraints~\cite{azure-zone,aws-zone,protean}) lead to correlated high demands being colocated in a rack, pooling could be less effective --- however, prior industry data has not observed such strong correlations~\cite{berger2023design}.


\subsection{PCIe Device Failures}





Many cloud servers use only a single NIC to keep costs low.
If this NIC fails, the entire server becomes unreachable and needs to wait for repair~\cite{lyu2023hyrax}.
Providers can design servers with redundant PCIe devices (e.g., NICs) to ensure availability during hardware failures.
However, PCIe devices are expensive and power-intensive.
Additionally, since device failures are not very common~\cite{lyu2023hyrax}, redundancy exacerbates stranding.

Pooling PCIe devices solves this problem by sharing redundant devices across multiple hosts.
If a device fails, the pool can allocate a replacement to the affected host.
This reduces the number of redundant devices needed, lowers costs, and improves device utilization.

\section{CXL Memory Pools}
\label{sec:cxl}

CXL standardizes interconnect protocols between processors and memory devices.
We focus on CXL.mem which enables cores and devices to transparent load/store (and DMA) to device memory~\cite{cxl-intro,cxl-3.0}.
CXL establishes a link based on the PCIe physical layer, ports, and cables.
It achieves low latency with custom link and transaction layers.
All major server CPUs (Intel, AMD, ARM) support CXL today.



\paragraph{Latency and bandwidth.}
The idle load-to-use latency of CXL memory is about 2-3$\times$ the latency of local DDR5 memory.
For example, recent measurements show 2.15$\times$ idle latency on a Leo CXL Smart Memory Controller ~\cite{cxl-intro}.
In terms of bandwidth, a CXL 2.0/PCIe-5.0 $\times$8 link matches the bandwidth of a DDR5-4800 channel at a typical 2:1 read to write ratio (30 GB/s).
CPUs can interleave at 256 Byte granularity across multiple CXL links to achieve higher bandwidth.
On Intel Xeon 6 (Granite Rapids), we can interleave across 64 CXL lanes per CPU socket~\cite{lenovo2024thinksystem,asrock2024gnrd8,kalodrich2024supermicro,kaytus2024kr2280v3, kennedy2024lenovo}, providing $\approx$ 240 GB/s.

\paragraph{CXL memory pools.} 
A CXL memory pool is a set of CXL memory devices that connect to multiple hosts, allowing them to dynamically allocate memory from the pool.
The set of hosts connected to a CXL pool is called a CXL pod~\cite{pasha}.

CXL memory pools can be constructed either by CXL devices with multiple ports or by CXL switches.
Devices with multiple CXL ports, or \emph{multi-headed devices (MHD)}, can connect to multiple hosts via dedicated CXL links~\cite{pond,ufx-news,astera-news,berger2025octopus,hyperscale2023cxl,marvell2024structera,ha2023dynamic,seagate2024cxl,mackey2024cxl}.
These devices are readily available today from AsteraLabs~\cite{leo2023}, Marvell~\cite{marvell2024structera}, and UnifabriX~\cite{ufx-news}.
A single MHD today has up to 20 CXL ports~\cite{ufx-news} and multiple MHDs can be combined to scale to larger CXL pod sizes~\cite{berger2025octopus}.

On the other hand,
CXL switches allow CXL devices (including devices with a single port) to be connected to a CXL switch, which in turn connects to multiple hosts.
Switched designs resemble tree network topologies in CXL 2.0~\cite{cxl2topology} and Clos network topologies in CXL 3.0~\cite{cxl3fabric}.
A single switch typically offers 128-256 CXL lanes~\cite{xconn-cxl}.
These lanes would be shared among connected CPUs, CXL controllers, and other devices.
For example, 128 lanes could connect six CPUs with x16 CXL ports and two single-ported CXL controllers with x16 ports.
In CXL 3.0, multiple switching levels can be used to build large-scale CXL pods with 4096 hosts~\cite{cxl-intro}.

The tradeoff for switched designs is higher power, cost, and latency.
Prior work observes that CXL switch costs can exceed the benefits that can be achieved from pooling at scale~\cite{berger2023design,against-cxl}.
CXL switch latency is also fundamentally higher than direct CXL connections, as switches require serialization and deserialization of CXL packets twice on every path from CPU to CXL memory controller~\cite{berger2023design, pond}.
Current switches add more than $250ns$ of latency~\cite{xconn2024latency}, resulting in idle load-to-use latency of roughly $500-600ns$.

In short, CXL switches are more costly and have higher latency overhead, so MHD-based pods will likely be deployed earlier~\cite{berger2023design,against-cxl,berger2025octopus}.

\paragraph{Shared CXL memory.} A CXL memory pool can be used as shared memory across all the hosts in a CXL pod.
Shared CXL memory has the potential to speed up RPCs, key-value stores, and distributed databases~\cite{hydrarpc,rpcool,ahn2024examination,cxl-shm,tigon,pasha,ctxnl,cxl-sync},

\paragraph{Cache coherence.} The CXL 3.0 specification introduces Back-Invalidate (BI), a hardware flow to implement cache coherence across multiple hosts~\cite{cxl-intro,cxl-3.0}.
With BI, a CXL 3.0 memory device can implement a snoop filter to track host caching and issue snoop requests to change the cache state of a host.
However, implementing BI involves both processor-side and device-side changes, which greatly increases hardware complexity and costs.
Neither CPUs nor CXL memory pool devices support BI today.


\section{Designing PCIe Pools in Software}
\label{sec:design}

\begin{figure*}[t]
    \centering
    \begin{subfigure}[t]{0.28\textwidth}
        \includegraphics[width=\columnwidth]{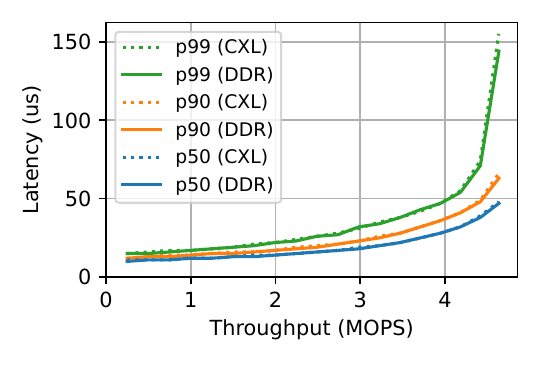}
        \vspace*{-7mm}
        \caption{75~B packets.}
        \label{fig:udp-75B}
    \end{subfigure}\hfill
    \begin{subfigure}[t]{0.28\textwidth}
        \includegraphics[width=\columnwidth]{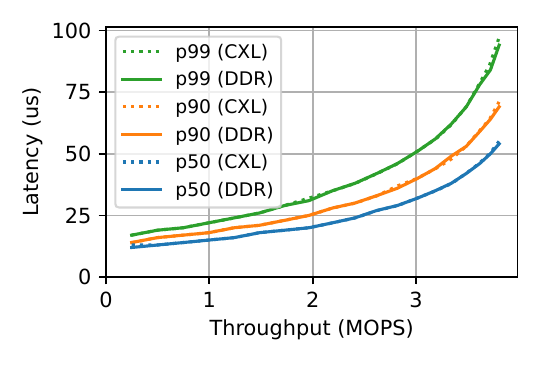}
        \vspace*{-7mm}
        \caption{1500~B packets.}
        \label{fig:udp-1500B}
    \end{subfigure}\hfill
    \begin{subfigure}[t]{0.28\textwidth}
        \includegraphics[width=\columnwidth]{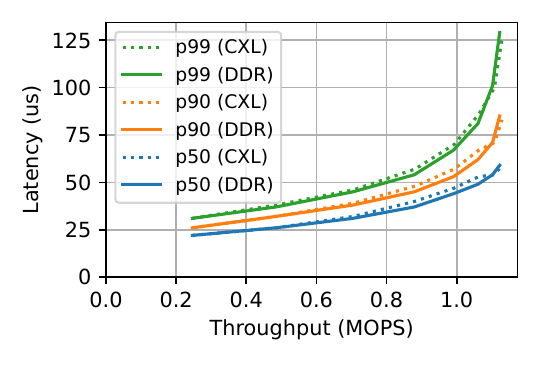}
        \vspace*{-7mm}
        \caption{9000~B packets.}
        \label{fig:udp-9000B}
    \end{subfigure}\hfill
    \vspace*{-2mm}
    \caption{Latency-throughput graph of the UDP microbenchmark with 100~Gbps NICs. On the server side, TX and RX buffers are allocated either from the CXL memory pool (dotted lines) or local DDR5 memory (solid lines).}
    \label{fig:udp}
\end{figure*}



We propose the following design goals to make our design practical and effective in unlocking the benefits of pooling:
\begin{denseenum}
    \item \textbf{Immediately Deployable.} 
    CXL memory pool devices available today do not have hardware cache coherence across hosts. Thus, pooling should work without coherence.
    \item \textbf{Device Compatibility.} Our design should be compatible with most PCIe devices (\eg NICs, SSDs, accelerators).
    \item \textbf{Load Shifting.} To realize load balancing and failovers, our design should support adjusting device-to-host mappings and shift loads between devices dynamically.
\end{denseenum}

Our design consists of two components: (i) \textbf{the datapath}, which
decouples PCIe devices from hosts by allowing hosts to access remote PCIe devices not directly connected to them via PCIe~(\S\ref{sec:datapath}); and (ii) \textbf{the pooling orchestrator}, which manages the mapping between PCIe devices and hosts
~(\S\ref{sec:orchestrator}).



Our design assumes containerized environments. The datapath is implemented in the host userspace I/O stacks, and the pooling orchestrator runs as a special management container on one of the hosts in the CXL pod.

Throughout this section, we use NICs as example PCIe devices as they exhibit lower latency and higher bandwidth compared to SSDs, making them more challenging to pool.
Nevertheless, our design is compatible with other PCIe devices, including SSDs and accelerators such as FPGAs.

\subsection{PCIe Datapath Over CXL}
\label{sec:datapath}
A CXL pool offers a shared memory range for connected hosts and their PCIe devices, which enables us to route PCIe traffic through the CXL pool across host boundaries.

\paragraph{Placing buffers in shared CXL memory} 
Routing PCIe traffic through a CXL memory pool requires placing I/O-related buffers in the CXL pool, so that remote PCIe devices can access these buffers via DMA.

For non-cache-coherent CXL pools, the datapath should explicitly maintain coherence in software. When modifying I/O buffers, the data should always be written to the CXL memory rather than staying in the CPU caches.
Otherwise, other hosts might retrieve stale data from the CXL memory.

\paragraph{Performance implications of CXL}
Although the latency overhead incurred by CXL could affect the I/O buffers placed in CXL memory pools, this overhead is negligible compared to the end-to-end I/O latency of NICs and SSDs.


To quantify if placing I/O buffers in CXL affects I/O latency, we set up a two-socket server equipped with a 100~Gbps Mellanox ConnectX-5 NIC and connect both of its CPUs to a MHD-based CXL pod with each a PCIe-5.0 $\times$8 link.
A second host serves as load generator (client) and connects via another 100~Gbps ConnectX-5 NIC to a common 100~Gbps switch.
We use Junction~\cite{junction}
as the network stack for both the server and the client.
We modify Junction on the server side to allocate TX and RX buffers (not the TX/RX queues) from the CXL memory pool.
The NIC connects to socket0 and uses one $\times$8 CXL link.
Junction runs on socket1 and uses the other $\times$8 CXL link.

We perform a UDP microbenchmark with varying payload sizes and compare the round-trip network latency of our modified Junction to an unmodified version that allocates memory locally and runs on socket0.
Figure~\ref{fig:udp} shows that, although CXL has higher access latency, placing TX/RX buffers in CXL has negligible effects on the network latency.
Maximum throughput is also not affected because the two PCIe-5.0 $\times$8 links provide enough bandwidth to saturate 100~Gbps NICs.
Scaling to two 400~Gbps NICs may be possible on a Xeon 6 CXL configuration (Section~\ref{sec:cxl}).
However, our proposal may work better for general-purpose computing than high-bandwidth HPC or ML use cases.




\begin{figure}
    \centering
    \includegraphics[width=0.85\columnwidth]{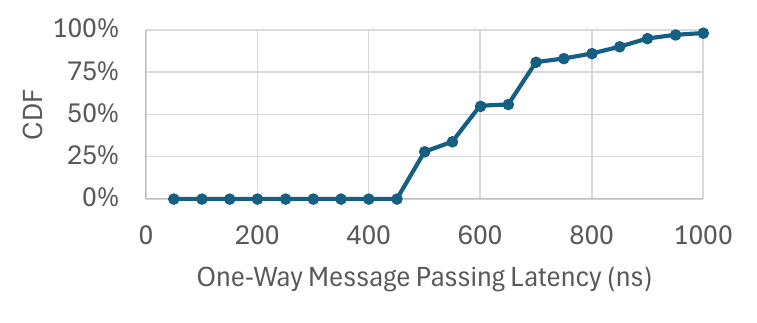}
    \vspace*{-6mm}
    \caption{Latency distribution of message passing.}
    \label{fig:msg-lat}
    \vspace*{-2mm}
\end{figure}

\paragraph{Event signaling and host-to-host communications}
Signaling events such as request arrivals often requires accessing PCIe device memory.
As a host cannot directly access a remote device's memory via MMIO, the datapath should provide a communication channel to forward MMIO operations to the host where the device is physically attached.



We prototype a shared-memory communication channel in shared CXL memory.
The channel is implemented as a ring buffer, with each message slot sized at 64~B to match the cacheline granularity.
It manages cache coherence in software by using non-temporal stores to send messages.

We measure its latency using a ping-pong test. The sender and receiver each connect to the CXL memory pool using a PCIe-5.0 $\times$16 link. 
Figure~\ref{fig:msg-lat} shows that shared-memory channels in CXL achieve sub-$\mu$s latencies without cache coherence. The median latency is around 600~ns, slightly above the theoretical minimum latency for message passing, which equals the total latency of one CXL write and one CXL read.



\subsection{Pooling Orchestrator}
\label{sec:orchestrator}

The pooling orchestrator manages a PCIe device pool.
It handles control plane operations, including allocating PCIe devices to hosts, monitoring resource usage and health status of each PCIe device, and migrating workloads between devices to balance load or handle device failures.

Each host runs a pooling agent that monitors and configures the PCIe device.
The orchestrator and the agents communicate using shared-memory channels in the shared CXL memory.
Both roles are good candidates to offload to accelerators such as SmartNICs.


When allocating a PCIe device to a host, the orchestrator first checks if the host has a local PCIe device that is below a load threshold.
If not, the orchestrator selects the least-utilized device in the pod to balance load.

If a PCIe device fails (\eg due to NIC link failures) or becomes overloaded, the corresponding agent will report the issue to the orchestrator using the shared-memory CXL channel. The orchestrator can then migrate workloads from the affected device to other devices.







\section{Discussion}
\label{sec:discussion}



The adoption of software-based PCIe pooling introduces numerous potential research opportunities.

\paragraph{Datacenter networks without ToRs}
Traditional datacenter designs have had a single top-of-rack (ToR) switch.
All servers in the rack connect to this ToR.
Links fan out from that ToR to multiple aggregation switches.
The ToR to aggregation links are typically oversubscribed~\cite{poutievski2022jupiter,singh2015jupiter,greenberg2009vl2}.

If a ToR fails, the entire rack becomes unreachable.
Datacenter operators have thus started to deploy dual ToRs, which avoids the single point of failure but increases cost~\cite{XieLu2025,qian2024alibaba}.

This creates an interesting opportunity: can we eliminate the ToRs altogether? Instead of oversubscribing at the ToR level, we can provision sufficient NICs within each CXL pod to provide equivalent oversubscription, and then directly connect these NICs to multiple switches within the aggregation layer.
This allows us to work around both ToR failures and NIC failures.
This would require high CXL pod reliability.

\paragraph{Highly-available CXL pods}
There are multiple ways to build CXL pods (Section~\ref{sec:cxl}).
MHD-based pods typically use multiple MHDs and thus inherently offer high redundancy.
A recent Microsoft white paper formalizes this with so-called dense topologies that offer $\lambda$ redundant paths within a CXL pool~\cite{berger2025octopus}.
Many industry proposals offer $\lambda=4$ or even $\lambda=8$~\cite{pond,ha2023dynamic,seagate2024cxl,hyatt2023quest}.

In the future, CXL 3.0 switches will explicitly support multi-path CXL topologies with Port Based Routing~\cite{cxl-intro}.

While CXL pod hardware may offer redundancy, correctly implementing redundancy and fail-over is a hard implementation challenge.
However, we believe that software implementations would likely still be significantly easier than hardware implementations based on PCIe switches.

\paragraph{Soft accelerator disaggregation}
The architecture commnuity has proposed and continues to propose an ever increasing set of accelerators.
Typically, they come in the form of PCIe cards~\cite{cong2014accelerator}.
Deploying an accelerator that is used by a large number of people (e.g., compression) is easy.
However, accelerators increasingly target highly specific workloads such as homomorphic encryption~\cite{zhang2024sok}.
If a user wants to use an accelerator, they need to run on a server with this accelerator.
Specialized accelerators may get infrequent use and thus may sit idle most of the time, which is a significant cost overhead.
Disaggregating accelerators through PCIe switches~\cite{gigaio-acc,gigaio-storage} is complex and expensive. Our approach allows exposing these accelerators as a disaggregated resource accessed via a CXL-connected pool.

\paragraph{PCIe pooling can complement RDMA-based storage disaggregation}
Storage disaggregation over RDMA is common in clouds.
However, in storage clusters, SSDs and NICs remain tightly coupled to hosts. Each host is provisioned with SSDs and NICs to handle peak demands.
Since storage clusters often exhibit skewed access patterns, NICs are mostly underutilized.
PCIe pooling can address this issue by pooling NICs between multiple storage nodes, improving resource efficiency in RDMA-based disaggregation.
 
\paragraph{Better host load balancing}
PCIe pooling can also facilitate load balancing for resources other than PCIe devices.
Typically, the challenge for moving load to another server is that TCP connections are assigned at setup and cannot be moved~\cite{hayakawa2021prism,choi2023capybara}.
A long-lived connection on a server whose load changes in the background may experience high latency.
Recent work has explored TCP connection migration to move requests to new servers after they are being processed, but this work requires programmable switches or other network middleboxes~\cite{hayakawa2021prism,choi2023capybara}.
Our virtual NIC approach could implement the transformations required to migrate connections seamlessly within the CXL pod.





\paragraph{CXL link bandwidth}
Recently platforms (\eg Intel Xeon 6) provide 64 CXL~2.0 / PCIe5 lanes per CPU socket~\cite{lenovo2024thinksystem,asrock2024gnrd8,kalodrich2024supermicro,kaytus2024kr2280v3, kennedy2024lenovo}.
Fully disaggregating a 200~Gbps NIC and a 400~Gbps NIC requires only 8 and 16 CXL lanes, respectively.
Therefore, pooling NICs to improve NIC utilization and reduce the NIC-to-host ratio can be supported by allocating enough CXL lanes per host to interleave between multiple MHDs.
Similarly, disaggregating six NVMe SSDs (current AWS servers offer six local NVMe SSDs~\cite{awsnetwork}, and datacenter SSDs today often provide 5~GB/s bandwidth~\cite{solidigmd5,990-EVO}) would require 30~GB/s bandwidth, which can be satisfied with 8 CXL lanes.

However, enabling a single host to saturate a large number of NICs to maximize its peak network performance does pose higher requirements for the CXL link~\cite{dfabric}.
For example, to drive the combined bandwidth of eight 400~Gbps NICs in a CXL pod, the host would need at least 100 CXL~2.0 lanes, making this use case less realistic.

\paragraph{Operational implications} 
Datacenters often need to update host OSes and platform firmware, which requires rebooting individual hosts.
To enable such maintenance, PCIe device pooling should support hot-adding and hot-removing hosts from a CXL pod, allowing updates to be rolled out incrementally on a per-host basis.
For instance, when reconfiguring a host in public clouds, the VM scheduler should live migrate all VMs off the host and notify the pooling orchestrator to hot-remove it.
Upon removal, the orchestrator should prevent new device allocations to the host and migrate its existing PCIe device assignments to other active hosts.

\section{Acknowledgments}

We thank the reviewers for their helpful comments. We also thank Rodrigo Fonseca and Stefan Saroiu for their feedback. This work was supported by NSF award CNS-2143868.
\label{lastpage}

\newpage

\bibliographystyle{plain}
\bibliography{database}

\end{document}